\documentclass[preprint,amsmath,amssymb,superscriptaddress]{revtex4-1}
\usepackage{graphicx}
\usepackage{dcolumn}
\usepackage{bm}
\usepackage{color}

\begin{document}

\title{Spin-polarized Weyl cones and giant anomalous Nernst effect in ferromagnetic Heusler films}

\author{Kazuki Sumida}
\email{sumida.kazuki@jaea.go.jp}
\affiliation{Materials Sciences Research Center, Japan Atomic Energy Agency, Hyogo 679-5148, Japan}
\affiliation{Department of Physics, Tokyo Institute of Technology, Tokyo 152-8551, Japan}
\affiliation{Graduate School of Science, Hiroshima University, 1-3-1 Kagamiyama, Higashi-Hiroshima 739-8526, Japan}
\author{Yuya Sakuraba}
\email{SAKURABA.Yuya@nims.go.jp}
\affiliation{Research Center for Magnetic and Spintronic Materials, National Institute for Materials Science, Sengen 1-2-1, Tsukuba 305-0047, Japan}
\affiliation{PRESTO, Japan Science and technology Agency, Saitama 332-0012, Japan}
\author{Keisuke Masuda}
\affiliation{Research Center for Magnetic and Spintronic Materials, National Institute for Materials Science, Sengen 1-2-1, Tsukuba 305-0047, Japan}
\author{Takashi Kono}
\affiliation{Graduate School of Science, Hiroshima University, 1-3-1 Kagamiyama, Higashi-Hiroshima 739-8526, Japan}
\author{Masaaki Kakoki}
\affiliation{Graduate School of Science, Hiroshima University, 1-3-1 Kagamiyama, Higashi-Hiroshima 739-8526, Japan}
\author{Kazuki Goto}
\affiliation{Research Center for Magnetic and Spintronic Materials, National Institute for Materials Science, Sengen 1-2-1, Tsukuba 305-0047, Japan}
\author{Weinan Zhou}
\affiliation{Research Center for Magnetic and Spintronic Materials, National Institute for Materials Science, Sengen 1-2-1, Tsukuba 305-0047, Japan}
\author{Koji Miyamoto}
\affiliation{Hiroshima Synchrotron Radiation Center, Hiroshima University, 2-313 Kagamiyama, Higashi-Hiroshima 739-0046, Japan}
\author{Yoshio Miura}
\affiliation{Research Center for Magnetic and Spintronic Materials, National Institute for Materials Science, Sengen 1-2-1, Tsukuba 305-0047, Japan}
\author{Taichi Okuda}
\affiliation{Hiroshima Synchrotron Radiation Center, Hiroshima University, 2-313 Kagamiyama, Higashi-Hiroshima 739-0046, Japan}
\author{Akio Kimura}
\email{akiok@hiroshima-u.ac.jp}
\affiliation{Graduate School of Science, Hiroshima University, 1-3-1 Kagamiyama, Higashi-Hiroshima 739-8526, Japan}
\affiliation{Graduate School of Advanced Science and Engineering, Hiroshima University, 1-3-1 Kagamiyama, Higashi-Hiroshima 739-8526, Japan}
\date{\today}

\begin{abstract}
Weyl semimetals are characterized by the presence of massless band dispersion in momentum space. When a Weyl semimetal meets magnetism, large anomalous transport properties emerge as a consequence of its topological nature. Here, using $in-situ$ spin- and angle-resolved photoelectron spectroscopy combined with $ab\ initio$ calculations, we visualize the spin-polarized Weyl cone and flat-band surface states of ferromagnetic Co$_2$MnGa films with full remanent magnetization. We demonstrate that the anomalous Hall and Nernst conductivities systematically grow when the magnetization-induced massive Weyl cone at a Lifshitz quantum critical point approaches the Fermi energy, until a high anomalous Nernst thermopower of $\sim 6.2$ $\rm \mu V K^{-1}$ is realized at room temperature. Given this topological quantum state and full remanent magnetization, Co$_2$MnGa films are promising for realizing high efficiency heat flux and magnetic field sensing devices operable at room temperature and zero-field.
\end{abstract}

\maketitle

\section{Introduction}
When electric and thermal currents flow through a ferromagnet, an electric field emerges orthogonally to the current path. The two effects are respectively called the anomalous Hall (AHE) and Nernst (ANE) effects and are exploited as operating mechanisms in various novel applications such as energy harvesting~\cite{2008Bell-Science, 2016Sakuraba-SM}, magnetic sensor~\cite{2011Vidal-APL}, and heat flux sensing~\cite{2020Zhou-APEX}. The associated transverse voltage of the electric field is empirically proportional to its spontaneous magnetization. In contrast to the general belief, recent discoveries of both large AHE and ANE, which do not scale with magnetization, have elicited great surprise~\cite{2015Nakatsuji-Nature, 2017Ikhlas-NatPhys, 2018Sakai-NatPhys, 2018Liu-NatPhys, 2019Guin-NPGAsia, 2019Guin-AdvMater}. In particular, the observed ANE thermopower of single crystalline bulk Co$_2$MnGa at room temperature, $\sim$6.0 $\rm \mu V K^{-1}$ is an order of magnitude larger than that of other ferromagnets with similar magnetizations~\cite{2018Sakai-NatPhys, 2019Guin-NPGAsia}. These transverse properties are postulated to arise from a Berry curvature emerging within band structures near the Fermi energy ($E_{\rm F}$)~\cite{2006Xiao-PRL, 2010Xiao-RMP}.

Topologically non-trivial Weyl semimetals possessing spin-split massless fermions characterized by zero-gap and linear band dispersions are promising candidates featuring a large Berry curvature~\cite{2011-WanPRB, 2015-SoluyanovNature, 2018Manna-PRX}. Weyl fermions in solids can be realized in materials that break inversion symmetry or time-reversal symmetry. With the breaking of such symmetries, Weyl nodes appear as pairs in momentum space and act as magnetic monopoles with positive and negative chirality~\cite{2017Yan-ARCMP}. To date, Weyl fermions have been verified in experiments in non-centrosymmetric (e.g., TaAs-family) and magnetic materials (e.g., Mn$_3$Sn) through angle-resolved photoelectron spectroscopy (ARPES) and magneto-transport measurements~\cite{2015Xu-Science, 2015Lv-NatPhys, 2015Lv-PRX, 2016Belopolski-NC, 2017Xu-SA, 2017Kuroda-NatMater, 2017Belopolski-NC, 2018Wang-NC, 2018Liu-NatPhys, 2019Borisenko-NC, 2019Liu-Science}.

Recently, a Co$_2$MnGa Heusler alloy has also been theoretically predicted to be a ferromagnetic Weyl semimetal with a high Curie temperature and has been experimentally demonstrated in the bulk form to exhibit large anomalous transport properties under an external magnetic field~\cite{2018Sakai-NatPhys, 2019Guin-NPGAsia, 2016Kuebler-EPL}. The nature of this highly symmetric crystal [Fig. \ref{fig_trans}(a)] creates mirror-symmetry-protected Weyl nodal lines in the band structure as encountered by theory and experiments~\cite{2017Chang-PRL, 2019Belopolski-Science}. However, the nodal lines lead to vanishing Berry curvature when integrated over the whole Brillouin zone~\cite{2019Noky-PRB, 2019Noky-AS} and cannot explain the observed phenomena. One way to obtain a large Berry curvature is to “gap out’’ their nodal lines using remanent magnetization or an external magnetic field (specifically, to break the mirror symmetry). Yet, the experimental evidence for broken mirror symmetry was not provided by the recent ARPES measurement on bulk Co$_2$MnGa crystal because the remanent magnetization was negligible as applying external magnetic fields is not permitted in this measurement.
For practical applications in which zero-field operation and gigantic outputs are a requirement, it is thus indispensable to truly understand the band structure responsible for the anomalous transport properties in films with full remanent magnetization.

Here, we experimentally and theoretically investigated the topological band structure and anomalous transport properties of ferromagnetic Co$_2$MnGa thin films. Growth of high-quality thin films possessing full remanent magnetization and {\it in-situ} spin-resolved ARPES (SARPES) measurements permit access to their non-trivial band structures modified by the broken mirror symmetry. We observed spin-polarized Weyl cones located mostly at a Lifshitz quantum critical point and a flat band of surface states. Furthermore, when the energy associated with the massive Weyl cone approaches $E_{\rm F}$, the anomalous Hall and Nernst conductivities systematically increase as the electron number rises. In particular, the ANE reaches thermopower of $\sim6.2$ $\rm \mu V K^{-1}$ at room temperature, which is the highest amongst magnetic films to the best of our knowledge.

\section{Results and Discussion}
\subsection{AHE and ANE properties of epitaxial Co$_2$MnGa films}

\begin{figure*}
\includegraphics[width=\columnwidth]{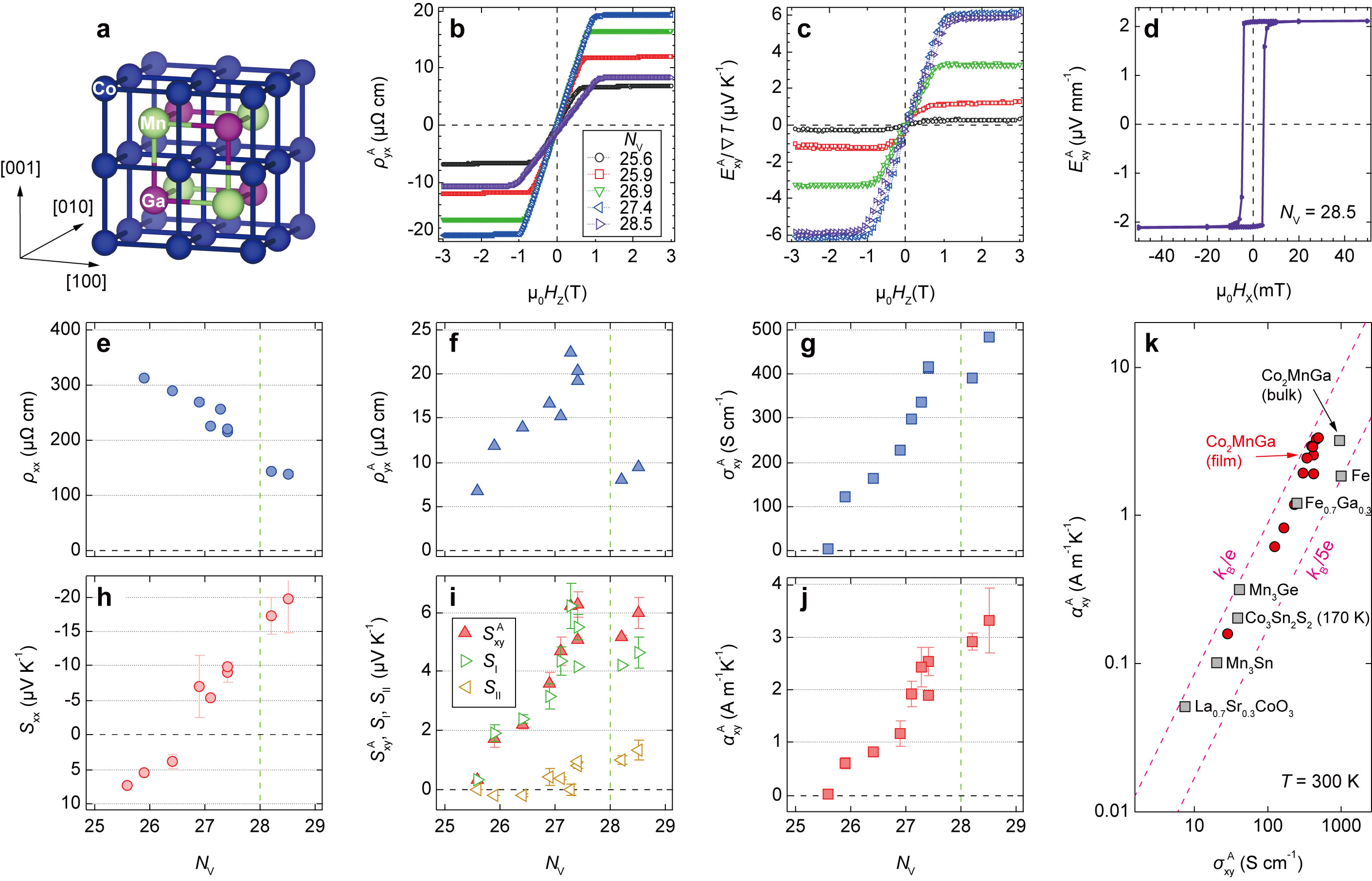}
\caption{\label{fig_trans} \bf {Anomalous transport properties of $L2_1$-ordered Co$_2$MnGa films.} \rm(a) Schematic of the $L2_1$ ordered structure of Co$_2$MnGa. Dependence on perpendicular external magnetic field ($H_{z}$) of (b) $\rho^{\rm A}_{yx}$ and (c) $E^{\rm A}_{xy}$ normalized by given temperature gradient $\nabla T_{x}$ in the Co$_2$MnGa films. (d) $E^{\rm A}_{xy}$ measured by giving the perpendicular temperature gradient $\nabla T_{z}$ and sweeping the in-plane magnetic field $H_{x}$ for the sample with $N_{\rm v}$ = 28.5. Dependence on total valence electron number $N_{\rm v}$ of (e) $\rho_{xx}$, (f) $\rho^{\rm A}_{yx}$, (g) $\sigma^{\rm A}_{xy}$, (h) $S_{xx}$, (i) $S^{\rm A}_{xy}$, $S_{\rm I}$ and $S_{\rm II}$ and (j) $\alpha^{\rm A}_{xy}$. Green dashed lines in (e--j) indicate $N_{\rm v}$ for the stoichiometric composition. All error bars represent standard deviation. (k) Relationship between $\alpha^{\rm A}_{xy}$ vs $\sigma^{\rm A}_{xy}$ in our Co$_2$MnGa films (red circles) and different magnets (gray squares: taken from Ref.~\cite{2020Xu-PRB}).}
\end{figure*}

To examine the influence of the relative location of the Weyl cone associated energy with respect to $E_{\rm F}$ on the transverse transport properties, we grew epitaxial Co$_2$MnGa films with different compositions and investigated their AHE and ANE properties. 
Highly $L2_1$-ordered structure was confirmed in all Co$_2$MnGa films by the quantitative analysis of X-ray diffraction patterns with taking the off-stoichiometry in each film into account (Supplementary Fig. 1 and the associated text).
Figures \ref{fig_trans}(b) and \ref{fig_trans}(c) show the dependence on the perpendicular external magnetic field $\rm \mu_0$$H_{z}$ of the anomalous Hall resistivity $\rho^{\rm A}_{yx}$ and the normalized transverse thermoelectric voltage $E^{\rm A}_{xy}$ for Co$_2$MnGa films with different compositions labelled H8, H7, H5, H1, and E2 [Table I], leading to a different number of valence electrons ($N_{\rm v}$) ranging from 25.6 (H8) to 28.5 (E2). These curves clearly show that for all these films the respective magnitude of the AHE and the ANE changes significantly depending on the composition ratio of Co$_2$MnGa regardless of its atomic ordering.
It should be mentioned that here we apply the $\nabla T_{x}$ and $H_{z}$ for evaluating the thermopower of the ANE strictly, causing no ANE signal at $\rm \mu_0$$H_{z}$ = 0 as the spontaneous magnetization does not appear to the $z$-direction due to the strong demagnetization field. 
Thus, we also measured ANE signal by applying the $\nabla T_{z}$ and sweeping the magnetic field in the plane and observed clear ANE voltage at zero-field due to the perfect spontaneous in-plane magnetization as shown
in Fig. \ref{fig_trans}(d), which is advantageous to thermoelectric applications such as heat flux sensor~\cite{2020Zhou-APEX}.

Resistivity $\rho^{\rm A}_{yx}$ and thermopower $S^{\rm A}_{xy}$ [Figs. \ref{fig_trans}(f) and \ref{fig_trans}(i)] were estimated by finding the intercept of the linear-fitted curve for the saturation region in Figs. \ref{fig_trans}(b) and \ref{fig_trans}(c), respectively. From Figs. \ref{fig_trans}(e) and \ref{fig_trans}(f), $\rho_{xx}$ monotonically decreases with increasing $N_{\rm v}$, whereas $\rho^{\rm A}_{yx}$ has a maximum value of 22.5 $\mu\Omega$cm at around $N_{\rm v}$ = 27.3 to 27.4. $\sigma^{\rm A}_{xy}$ evaluated from $\sigma^{\rm A}_{xy}$=$\rho^{\rm A}_{yx}$/($\rho_{xx}^2$ + ${\rho^{\rm A}_{yx}}^2$) exhibits a nearly monotonic increase from 2 $\rm S cm^{-1}$ at $N_{\rm v}$ = 25.6 to 485 $\rm S cm^{-1}$ at 28.5 [Fig. \ref{fig_trans}(g)].

Figure \ref{fig_trans}(h) displays the $N_{\rm v}$ dependence of the Seebeck coefficient $S_{xx}$. In the small $N_{\rm v}$ region, $S_{xx}$ is positive and gradually decreases with $N_{\rm v}$ changing sign at $N_{\rm v}$ = 26.7. A very tiny ANE is observed in the small $N_{\rm v}$ region [Fig. \ref{fig_trans}(i)]; for instance, 0.3 $\rm \mu V K^{-1}$ at $N_{\rm v}$ = 25.6. 
Interestingly, $S^{\rm A}_{xy}$ increases steeply with increasing $N_{\rm v}$ with the largest $S^{\rm A}_{xy}$ achieved being 6.2 $\rm \mu V K^{-1}$ at $N_{\rm v}$ = 27.3, which is the highest in the ferromagnetic thin films~\cite{2018Reichlova-APL, 2020Park-PRB} and slightly smaller than the highest value in bulk Co$_2$MnGa $\sim$8.0 $\rm \mu V K^{-1}$~\cite{2020Xu-PRB}.
$S^{\rm A}_{xy}$ is obtained from the linear response equation,
\begin{equation}
S^{\rm A}_{xy}=\rho_{xx} \alpha^{\rm A}_{xy} - \rho^{\rm A}_{yx} \alpha_{xx}
\end{equation}
Here, $\alpha_{xx}$ and $\alpha^{\rm A}_{xy}$ are the longitudinal and transverse thermoelectric conductivities, respectively.
This equation indicates that there are two different phenomenological contributions to ANE. To simplify our explanation, we denote the associated terms as $S_{\rm I}$ = $\rho_{xx} \alpha^{\rm A}_{xy}$ and $S_{\rm II}$ = $- \rho^{\rm A}_{yx} \alpha_{xx}$. As mentioned in Ref.~\cite{2019Nakayama-PRM, 2020Sakuraba-PRB}, $S_{\rm II}$ is regarded as the contribution of AHE to ANE induced by a Seebeck-driven longitudinal current. Similarly, $S_{\rm I}$ stems from the direct conversion from a temperature gradient to a transverse current via $\alpha^{\rm A}_{xy}$. $S_{\rm II}$ can be converted to $-S_{xx}\rho^{\rm A}_{yx}/\rho_{xx}$, enabling a direct estimate from experimental data plotted in Fig. \ref{fig_trans}(i). One clearly views the contribution of $S_{\rm II}$ to $S^{\rm A}_{xy}$ to be very limited. Therefore, $S_{\rm I}$ $(= S^{\rm A}_{xy} - S_{\rm II})$ dominates the observed $S^{\rm A}_{xy}$ over the whole range of $N_{\rm v}$. Transverse thermoelectric conductivity $\alpha^{\rm A}_{xy}$ estimated from $S_{\rm I}$ and $\rho_{xx}$ increases enormously with $N_{\rm v}$ in a similar way to $\sigma^{\rm A}_{xy}$ and has a maximal value of 3.3 $\rm A m^{-1}K^{-1}$ at $N_{\rm v}$ = 28.5 [Fig. \ref{fig_trans}(j)].
Figure \ref{fig_trans}(k) shows the relationship between $\alpha^{\rm A}_{xy}$ and $\sigma^{\rm A}_{xy}$ at 300 K for Co$_2$MnGa films compared to other magnets.
A previous study by Xu $et\ al$.~\cite{2020Xu-PRB} clarified that many topological magnets have a universal relationship $\alpha^{\rm A}_{xy}/\sigma^{\rm A}_{xy} \sim k_{\rm B}/e$.
Thus, we examined this ratio in our present samples as shown in Fig. \ref{fig_trans}(k).
While both $\alpha^{\rm A}_{xy}$ and $\sigma^{\rm A}_{xy}$ are enhanced more than one order of magnitude by increasing $N_{\rm v}$ from 25.6 to 28.5, we confirmed that the $\alpha^{\rm A}_{xy}/\sigma^{\rm A}_{xy}$ ratios of our Co$_2$MnGa films also follow this universal behavior.
It signifies that the main origin of AHE and ANE in all Co$_2$MnGa films comes from the intrinsic topological nature~\cite{2020Xu-PRB}.

We also found that $\rho_{xx}$ decreases with increasing $N_{\rm v}$ whereas $S^{\rm A}_{xy}$ increases in our Co$_2$MnGa films [Figs. \ref{fig_trans}(e) and \ref{fig_trans}(i)].
This relation between $\rho_{xx}$ and $S^{\rm A}_{xy}$ is opposite to that reported in Co$_3$Sn$_2$S$_2$, where $S^{\rm A}_{xy}$ is enhanced by increasing $\rho_{xx}$~\cite{2019Ding-PRX}.
It is worth mentioning that this discrepancy arises from a difference in an origin of enlargement of $S^{\rm A}_{xy}$.
Namely, Ding $et\ al$. prepared Co$_3$Sn$_2$S$_2$ single crystals having different impurity concentrations and found that the impurity scatterings in Co$_3$Sn$_2$S$_2$ enlarge $\rho_{xx}$ but preserve $\alpha^{\rm A}_{xy}$, resulting in the enlargement of $S^{\rm A}_{xy}$ through $S_{\rm I}$ term.
On the other hand, this study tunes the position of $E_{\rm F}$ by adjusting the composition of Co$_2$MnGa film, leading to a drastic enlargement of $\alpha^{\rm A}_{xy}$ with $N_{\rm v}$ which is large enough to overcome the reduction of $\rho_{xx}$, thus $S^{\rm A}_{xy}$ is enhanced through $S_{\rm I}$ term as well.

\subsection{Theoretical calculations}

\begin{figure*}
\includegraphics[width=\columnwidth]{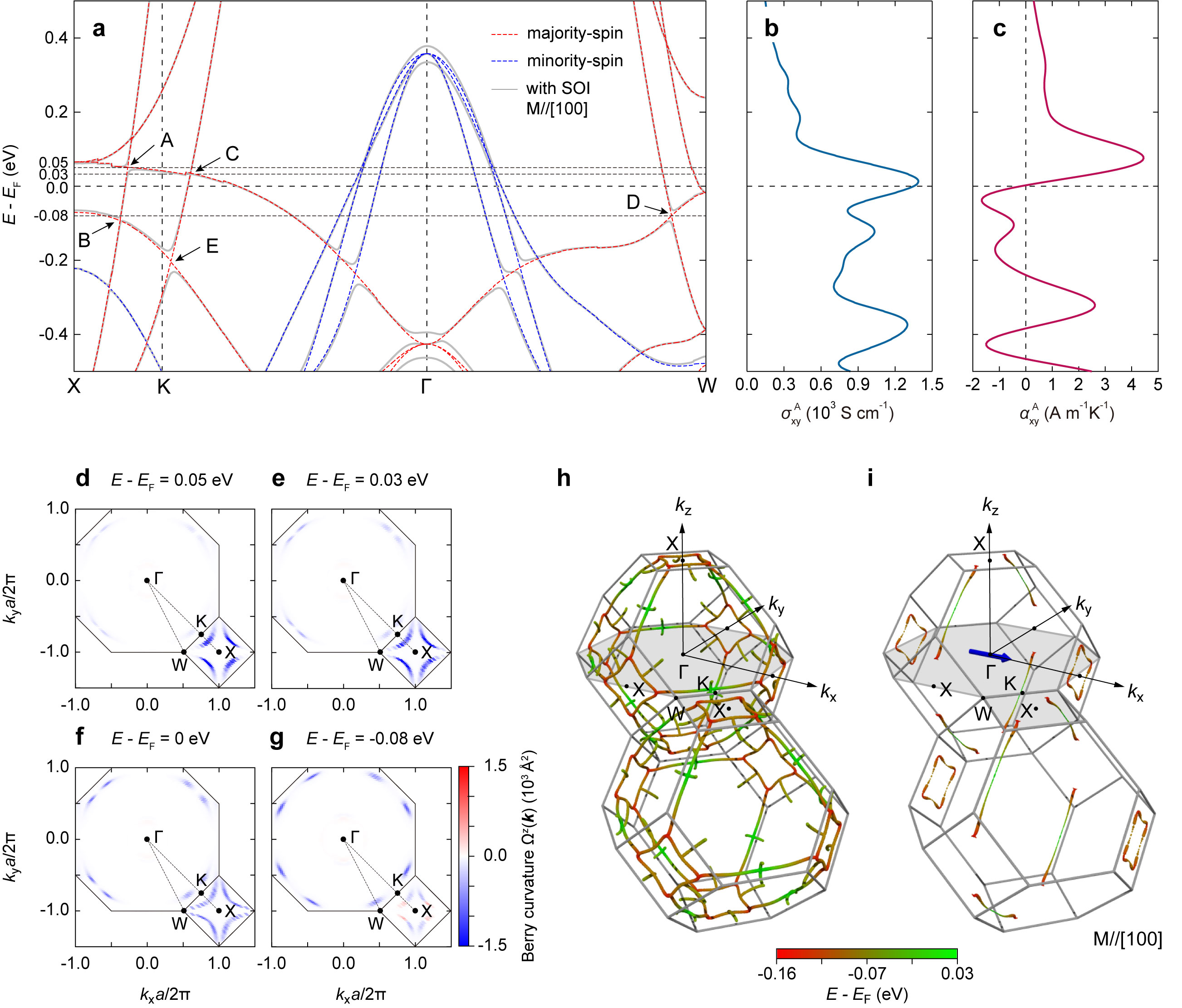}
\caption{\label{fig_theory} \bf {Calculated band structure, Berry curvature, and anomalous Hall and Nernst conductivities of Co$_2$MnGa.} \rm(a) Band structures and calculated (b) $\sigma^{\rm A}_{xy}$ and (c) $\alpha^{\rm A}_{xy}$ of $L2_1$-ordered Co$_2$MnGa. (d)--(g) Berry curvatures $\Omega^{z}({\bf k})$ calculated at $E-E_{\rm F}=0.05$, 0.03, 0, and $-0.08\,{\rm eV}$, respectively. (h) Three-dimensional view of the Weyl nodal lines in the first and second Brillouin zones of Co$_2$MnGa, where the $k_z=0$ and $k_z=2\pi/a$ planes are highlighted, in the absence of SOI. Colour corresponds to the location of the Weyl node. (i) Same as (h) but with SOI and magnetization along [100]. The blue arrow indicates the direction of magnetization.}
\end{figure*}

To understand the origin of the strong $N_{\rm v}$ dependence of AHE and ANE, we have performed {\it ab initio} calculations. Figure \ref{fig_theory}(a) shows the band structure of Co$_2$MnGa along the X--K line at $k_z=2\pi/a$ and along the K--$\Gamma$--W line at $k_z=0$ plane in the Brillouin zone [Fig. \ref{fig_theory}(h)]. The red and blue dashed lines represent the majority- and minority-spin bands, respectively, in the absence of the spin--orbit interaction (SOI). A large minority-spin hole pocket is found around $\Gamma$, whereas majority-spin bands dominate near $E_{\rm F}$ around X, K, and W and form crossing bands, labelled A through E. Previous studies have shown that such majority-spin band structures have three types of Weyl nodal loops in the Brillouin zone reflecting mirror symmetries with respect to the $k_i=0$ planes ($i=x,y,z$)~\cite{2016Wang-PRL,2018Noky-PRB,2017Chang-PRL, 2018Manna-PRX,2019Guin-NPGAsia}. For example, the nodal points of bands C and D in Fig. \ref{fig_theory}(a) are connected by the nodal loop shown in Fig. \ref{fig_theory}(h). 

When the SOI and the magnetization along the [100] direction are taken into account, only the mirror plane $k_x=0$ remains; the other planes disappear. This is because the mirror plane perpendicular to the magnetization conserves the direction of spins whereas that parallel to the magnetization does not~\cite{2016Wang-PRL,2018Noky-PRB,2018Manna-PRX,2019Guin-NPGAsia}. Once the SOI is present, the energy gaps open at all points of the Weyl cones A to E, since the mirror symmetry is broken with respect to the $k_z=0$ and the $k_z=2\pi/a$ planes [Fig. \ref{fig_theory}(a), grey curves]. Specifically, several Weyl nodal loops collapse and become massive (gapped) Weyl cones, and only those protected by $k_x=0$ mirror symmetry survive [Fig. \ref{fig_theory}(i) and Supplementary Fig. 7].
Here, we emphasize that the gapped Weyl cones A and C are tilted, most being at a Lifshitz quantum critical point between type-I and type-II Weyl fermions.

From the calculated anomalous Hall conductivity $\sigma^{\rm A}_{xy}$ [Figure \ref{fig_theory}(b)], we see that $\sigma^{\rm A}_{xy}$ has large values of $\sim 10^3\, {\rm S cm^{-1}}$ and exhibit a peak near $E_{\rm F}$, consistent with previous results~\cite{2018Sakai-NatPhys, 2019Guin-NPGAsia}. In Fig. \ref{fig_theory}(c), we show the calculated transverse thermoelectric conductivity $\alpha^{\rm A}_{xy}$. Around $E=E_{\rm F}$, $\alpha^{\rm A}_{xy}$ increases with increasing $E-E_{\rm F}$ (i.e., by electron doping) and takes a maximum at $E-E_{\rm F}=0.07\,{\rm eV}$, near to where $\sigma^{\rm A}_{xy}$ sharply drops [Fig. \ref{fig_theory}(b)]. This is reasonable as $\alpha^{\rm A}_{xy}$ is approximately proportional to $-d\sigma^{\rm A}_{xy}/dE$ (Methods, Eq. (\ref{eq_alphaxy}), and the Sommerfeld expansion \cite{Ashcroft-Mermin}).

To discuss the correlation between the large $\sigma^{\rm A}_{xy}$ around $E=E_{\rm F}$ [Fig. \ref{fig_theory}(b)] and the band structures [Fig. \ref{fig_theory}(a)], we plotted the $(k_x,k_y)$ dependences of the Berry curvature $\Omega^{z}({\bf k})$ [Fig. \ref{fig_theory}(d)--(g)]; the nodal loops at the $k_z=0$ and $2\pi/a$ planes are marked as highlighted areas in Fig. \ref{fig_theory}(i). At $E-E_{\rm F}$ = 0.05 and 0.03 eV [Fig. \ref{fig_theory}(d) and (e)], the Berry curvature has large values on the X--K line but has small values on the K--$\Gamma$--W line, because the gap of Weyl cone A mainly contributes to the Berry curvature at these energies [Fig. \ref{fig_theory}(a)]. In contrast, at the lower energy of $E-E_{\rm F}=-0.08\,{\rm eV}$ [Fig. \ref{fig_theory}(g)], large values of the Berry curvature are obtained close to the W point on the $\Gamma$--W line, which is determined by the gap of Weyl cone D [Fig. \ref{fig_theory}(a)].
From these results, we conclude that both gaps open on two different nodal loops in the $k_z=0$ and $k_z=2\pi/a$ planes and yield a large Berry curvature at $E=E_{\rm F}$ [Fig. \ref{fig_theory}(f)], leading to a large anomalous Hall conductivity. The calculated $\sigma^{\rm A}_{xy}$ and $\alpha^{\rm A}_{xy}$ qualitatively explain the experimental results [Fig. \ref{fig_trans}(g) and (j)]. Specifically, the electron-doped sample exhibits large anomalous transport properties. However, the simple $E_{\rm F}$ shift that follows the rigid band model based on the stoichiometric Co$_2$MnGa gives a quantitative discrepancy with the $N_{\rm v}$ dependence of $\sigma^{\rm A}_{xy}$ and $\alpha^{\rm A}_{xy}$ obtained from experiments. For example, the calculated negative $\alpha^{\rm A}_{xy}$ in the $E - E_{\rm F} < 0$ region is not observed in the experimentally hole-doped samples. This discrepancy may be caused by an extrinsic mechanism or the formation of anti-site defects arising from off-stoichiometric compositions, which is not taken into account in the calculations.

\subsection{Band structures of an electron-doped Co$_2$MnGa film}

\begin{figure*}
\includegraphics[width=\columnwidth]{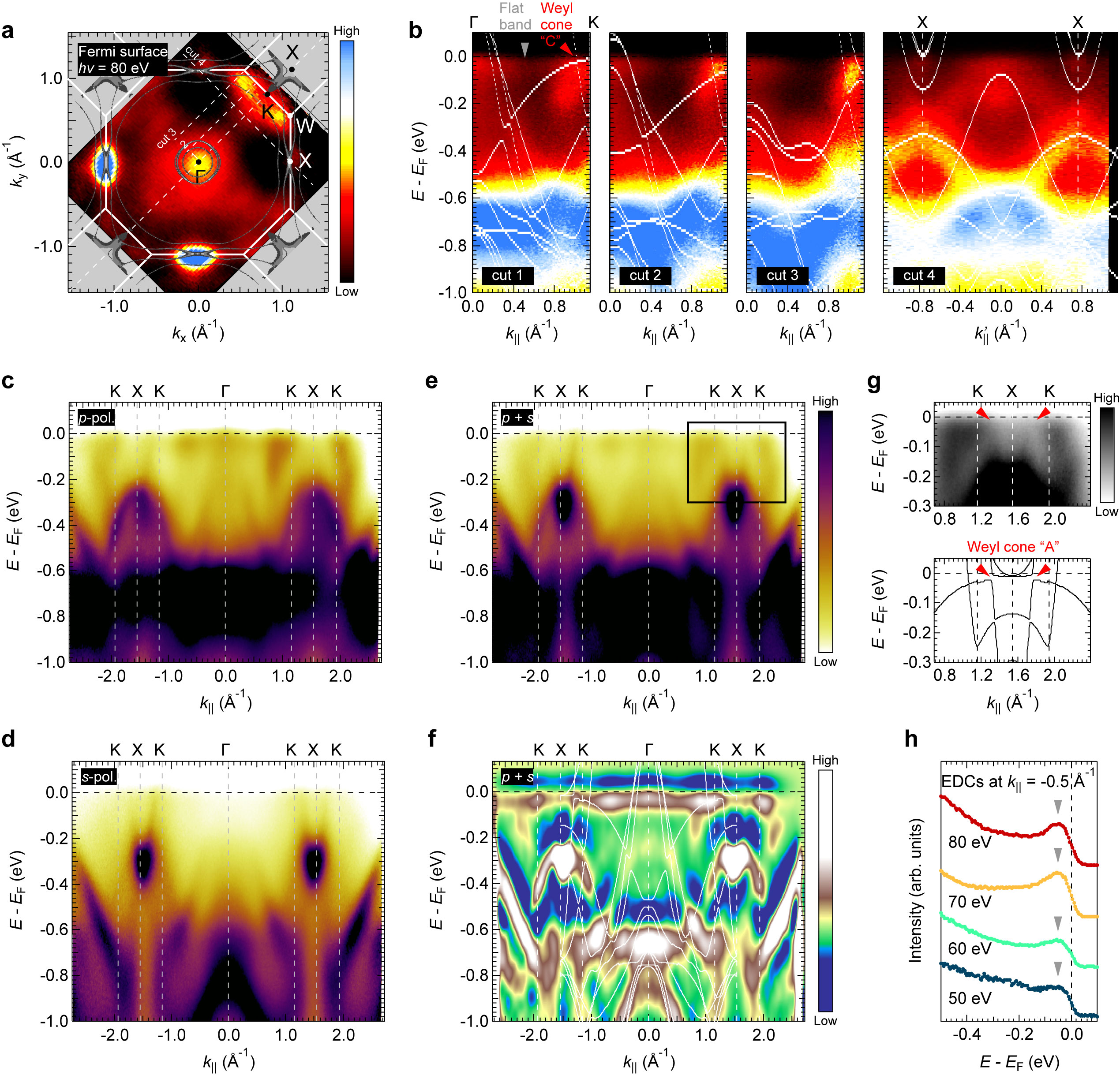}
\caption{\label{fig_ARPES} \bf {Observed band structures of an electron-doped Co$_2$MnGa film.} \rm(a) Fermi surface of sample E2 ($N_{\rm v}=28.5$) obtained at 80 eV with $p$-polarized light. The calculated Fermi surface of the stoichiometric Co$_2$MnGa ($N_{\rm v}=28.0$) with 70 meV chemical potential shifted is overlaid. (b) Observed and calculated band structures along cuts 1 to 4 marked by white dashed lines in (a). (c, d) Wide ARPES images along the $\Gamma$--K--X line recorded at 80 eV with $p$- and $s$-polarizations, respectively. (e) ARPES image with $p$-polarized light (c) is added with that with $s$-polarized light (d). (f) Second derivative ARPES image of (e) with respect to the energy and momentum directions. The calculated band dispersion is overlaid. (g) Magnified ARPES image (upper) in the frame in (e) and corresponding calculation (lower). (h) Photon-energy dependent EDCs at $k_{||}$ = $-$0.5 $\rm \AA^{-1}$ taken at 50, 60, 70, and 80 eV with $p$-polarized light.}
\end{figure*}

We performed ARPES experiments for the E2 ($N_{\rm v}=28.5$) sample as it exhibits the highest $\sigma^{\rm A}_{xy}$ and $\alpha^{\rm }_{xy}$, to determine the $E_{\rm F}$ location and the band structure with the mirror-symmetry breaking that yields a large Berry curvature as well as anomalous transport properties. Figure \ref{fig_ARPES}(a) shows the observed Fermi surface recorded at 80 eV photon energy after magnetization along [100] (Supplementary Movie 1 for a detailed continuous change in constant energy maps). Around the $\Gamma$ point, a circular shaped contour is evident. At each X point of the first Brillouin zone, we recognize a point-like structure. The calculated constant energy contour of the stoichiometric Co$_2$MnGa ($N_{\rm v}=28.0$) with the SOI along [100] was also plotted [Fig. \ref{fig_ARPES}(a)]. Except for the intensities in between $\Gamma$ and K, the features observed in the experiments are well reproduced by the calculation when the chemical potential shifts by +70 meV (Supplementary Fig. 5 and the associated text). Figure \ref{fig_ARPES}(b) shows the ARPES images and the band dispersions calculated at several momentum cuts [Fig. \ref{fig_ARPES}(a), white dashed lines]. At cut 1 ($\Gamma$-K line), we confirmed the large hole-pocket crossing $E_{\rm F}$ around $\Gamma$. One also finds distinct features that get closer to the K points with increasing $E - E_{\rm F}$ and finally cross $E_{\rm F}$, indicated by a red inverted triangle. These features are consistent with the calculated minority-spin hole band at $\Gamma$ point and one branch of the tilted majority-spin Weyl cone C [Fig. \ref{fig_theory}(a)]. In going from cut 1 to cut 3, the slope of the Weyl cone evolves to be sharper for both experiment and calculation. Here we also emphasize that an almost non-dispersive band is observed just below $E_{\rm F}$ around the $\Gamma$ point represented by a grey inverted triangle at cut 1. This flat band corresponds to the Fermi surface between the $\Gamma$ and K points [Fig. \ref{fig_ARPES}(a)] and is not reproduced by the calculations. At the cut 4 (X-X line), we can confirm the hole bands with energy maximum of $E - E_{\rm F}$ $\sim -100$ and $-$350 meV around $k'_{||}$ = 0 and $\pm0.8$ $\rm \AA^{-1}$, respectively, in both experiment and calculation. We also realize that the bottom of the electron bands located near $E_{\rm F}$ at the X points create prominent point-like structures on the Fermi surface [Fig. \ref{fig_ARPES}(a)]. 

Figure \ref{fig_ARPES}(c) and (d) shows the wide range ARPES images along the $\Gamma$--K--X line (cut 1) taken at 80 eV energy for incident photons with $p$- and $s$-polarization, respectively. With $s$-polarized light, a sharp electron-pocket is markedly enhanced around the X points, whereas the photoemission intensities of the tilted Weyl cone C and the flat band observed by $p$-polarization have mostly diminished. In Fig. \ref{fig_ARPES}(e) and (f), we present the ARPES image and its second derivative with the calculated band structure. Here, to eliminate the effect of the light polarization-dependent matrix-element, these images acquired with $p$- and $s$-polarized light are mixed. The experimental result is well reproduced by the calculations with $E_{\rm F}$ shifted upward by 70 meV (i.e., electron doping). This chemical potential shift is consistent with a higher $N_{\rm v}$ of 28.5 for this sample than the stoichiometric one ($N_{\rm v}=28.0$). Small discrepancies are noted between the observed and calculated band dispersions [Fig. \ref{fig_ARPES}(f)], for instance, the location of the bottom of the band of the sharp electron-pocket around the X points. The differences may arise through correlation or $k_z$ broadening effects~\cite{2019Nawa-RSC, 2003Strocov-JESRP}. 

Figure \ref{fig_ARPES}(g) shows the magnified ARPES image in the frame shown in Fig. \ref{fig_ARPES}(e). With suppression through the matrix-element effect, the band structure around the X point at the $k_z=2\pi/a$ plane in the second Brillouin zone are clearly visualized. In a comparison with the calculation [Fig. \ref{fig_ARPES}(g), lower panel], we realized that the observed band structure around the X point resembles the tilted and gapped Weyl cone A [Fig. \ref{fig_theory}(a)]. Because the upper part of the Weyl cone cannot be seen, $E_{\rm F}$ is probably located in the gap of the massive Weyl cone A.

Here, we turn our attention to the flat band observed around the $\Gamma$ point. To clarify the origin of the flat band, we show the photon-energy dependent energy distribution curves (EDCs) at $k_{||}$ = $-$0.5 $\rm \AA^{-1}$ taken after magnetization [Fig. \ref{fig_ARPES}(h)]. The prominent peaks caused by the flat band do not show a clear photon-energy ($k_{z}$) dependence. Details of the photon energy dependence can be find in Supplementary Fig. 6 and the associated text. We therefore conclude that the observed flat band just below $E_{\rm F}$ belongs to a surface state.

\begin{figure*}
\includegraphics[width=\columnwidth]{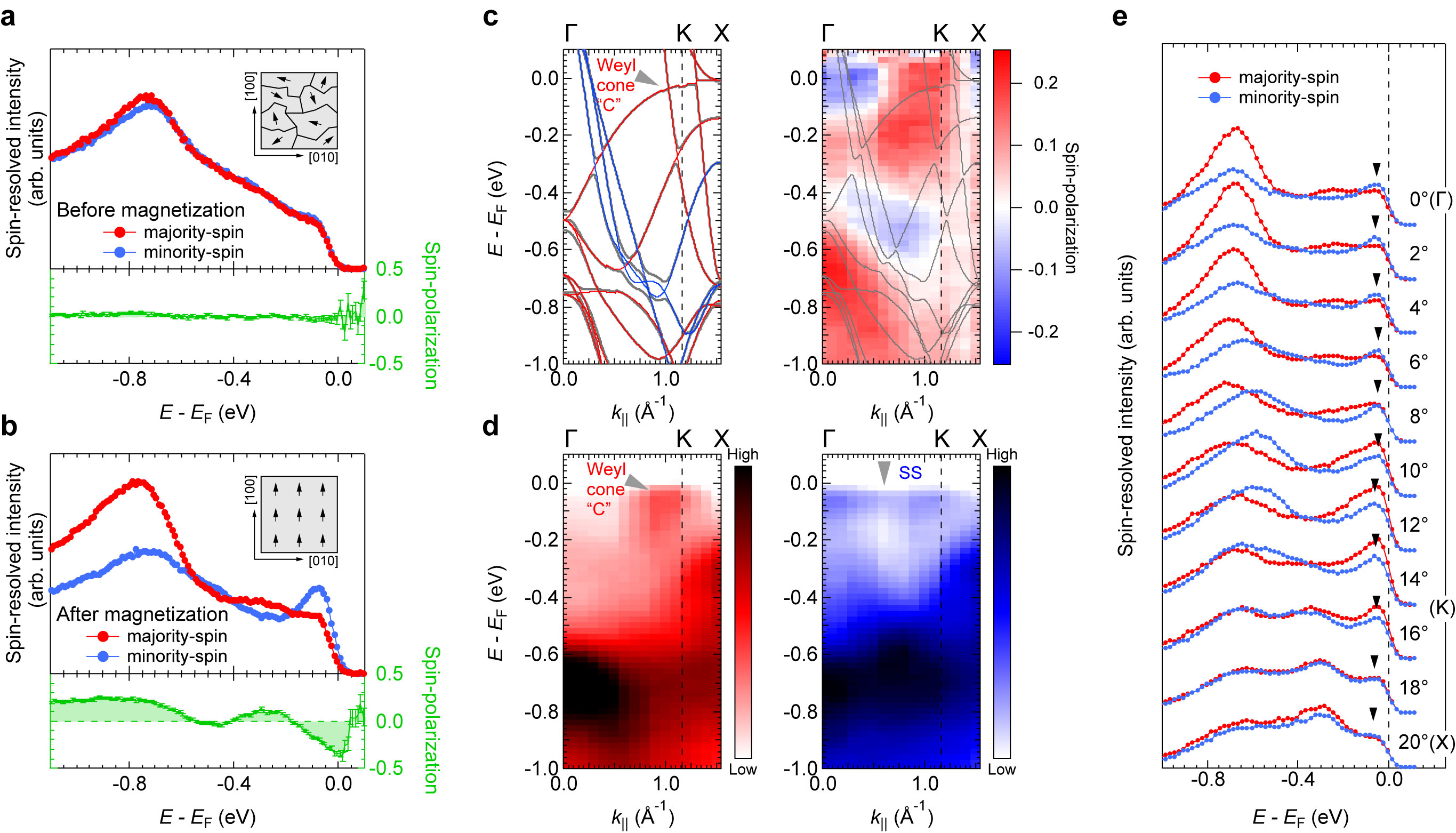}
\caption{\label{fig_spin} \bf {Spin-polarized Weyl cone and flat band surface state of Co$_2$MnGa film.} \rm(a,b) Spin-resolved EDCs and spin-polarizations at $\theta_{||}$ = $-$4$^\circ$ ($k_{||}$ $\sim$ $-$0.3 $\rm \AA^{-1}$) taken before and after magnetization at 80 eV with $p$-polarized light. All error bars represent standard deviation. The insets show the schematics of multi-magnetic domains and a single-magnetic domain. (c) Calculated band dispersion with $E_{\rm F}$ shifted upward by 70 meV (left) and spin-polarization map along the $\Gamma$--K--X line (right). (d) Spin-resolved ARPES images of the majority (left) and minority (right) spin states. (e) Spin-resolved EDCs taken from $\theta_{||}$ = 0$^\circ$ to 20$^\circ$ ($\Gamma$--K--X). The background signal of the secondary electrons was subtracted. Inverted triangles indicate the peak position of the minority spin-polarized flat band.}
\end{figure*}

To gain deeper insight into the Weyl fermion and the peculiar surface state in the Co$_2$MnGa film, we performed spin-resolved measurements. 
Since the light spot size ($>$1 mm) is much larger than the magnetic domain, we carried out SARPES measurements before and after magnetizing the sample.
Figure \ref{fig_spin}(a) and (b) shows the spin-resolved EDCs and spin-polarizations at $\theta_{||}$ = $-$4$^\circ$ ($k_{||}$ $\sim$ $-$0.3 $\rm \AA^{-1}$) taken before and after magnetizations, respectively. 
Although the spin-polarization is negligible over the whole energy region before magnetization due to the formation of magnetic domains, it is enhanced enormously after magnetization. In particularly, the large negative spin-polarization ($\sim$40 \%) originating from the flat surface state has been observed at $E_{\rm F}$. Figure \ref{fig_spin}(c) shows the calculated band structures (left) and the experimentally observed spin-polarization map along the $\Gamma$--K--X line overlaid with the calculated band structures (right). Here the positive (negative) spin-polarizations are marked in red (blue). Also, the spin-resolved band structures in the majority (left) and minority (right) spin channels [Figure \ref{fig_spin}(d)] feature a large hole band around the $\Gamma$ point having a minority-spin component. There are bands having a strong majority-spin component around the K point near $E_{\rm F}$. From calculations, the Weyl cone C is characterized by the majority-spin component and tilting [see Fig. \ref{fig_theory}(a)]. Furthermore, because Weyl cones A and C are mostly at Lifshitz quantum critical points, they have a large density of states at $E_{\rm F}$ when the Weyl node approaches $E_{\rm F}$~\cite{2018Sakai-NatPhys}. These features are confirmed by our ARPES and SARPES measurements, and therefore we conclude that the observed spin-polarized feature with positive spin-polarization near the K point can be ascribed to one flatter branch of the tilted spin-polarized Weyl cone C induced by broken mirror symmetry.

In addition, we find that the flat surface state has a minority-spin component [Fig. \ref{fig_spin}(d), right], the sign of which is opposite to that of the Weyl cone [Fig. \ref{fig_spin}(d), left]. Figure \ref{fig_spin}(e) shows the spin-resolved EDCs taken from $\theta_{||}$ = 0$^\circ$ to 20$^\circ$, which corresponds to the $k$-line from $\Gamma$ to X. The clear minority-spin peaks persist over a wide momentum region marked with inverted triangles. There are two main possibilities for the origin of this peculiar surface state. One is the topologically non-trivial Fermi-arc surface state~\cite{2015Xu-Science, 2015Lv-NatPhys, 2015Lv-PRX, 2016Belopolski-NC, 2017Belopolski-NC, 2019Borisenko-NC, 2019Liu-Science}. The other is a trivial surface resonance state, which was predicted for half-metallic Co-based Heusler alloys~\cite{2012Wustenberg-PRB, 2015Braun-PRB}. Having considered the location of the Weyl node before magnetization, it seems that the minority-spin surface state connects the Weyl cones at positive and negative momenta although further study is needed to elucidate the origin of the surface state.

\subsection{Band structures of a hole-doped Co$_2$MnGa film}

\begin{figure*}
\includegraphics[width=12cm]{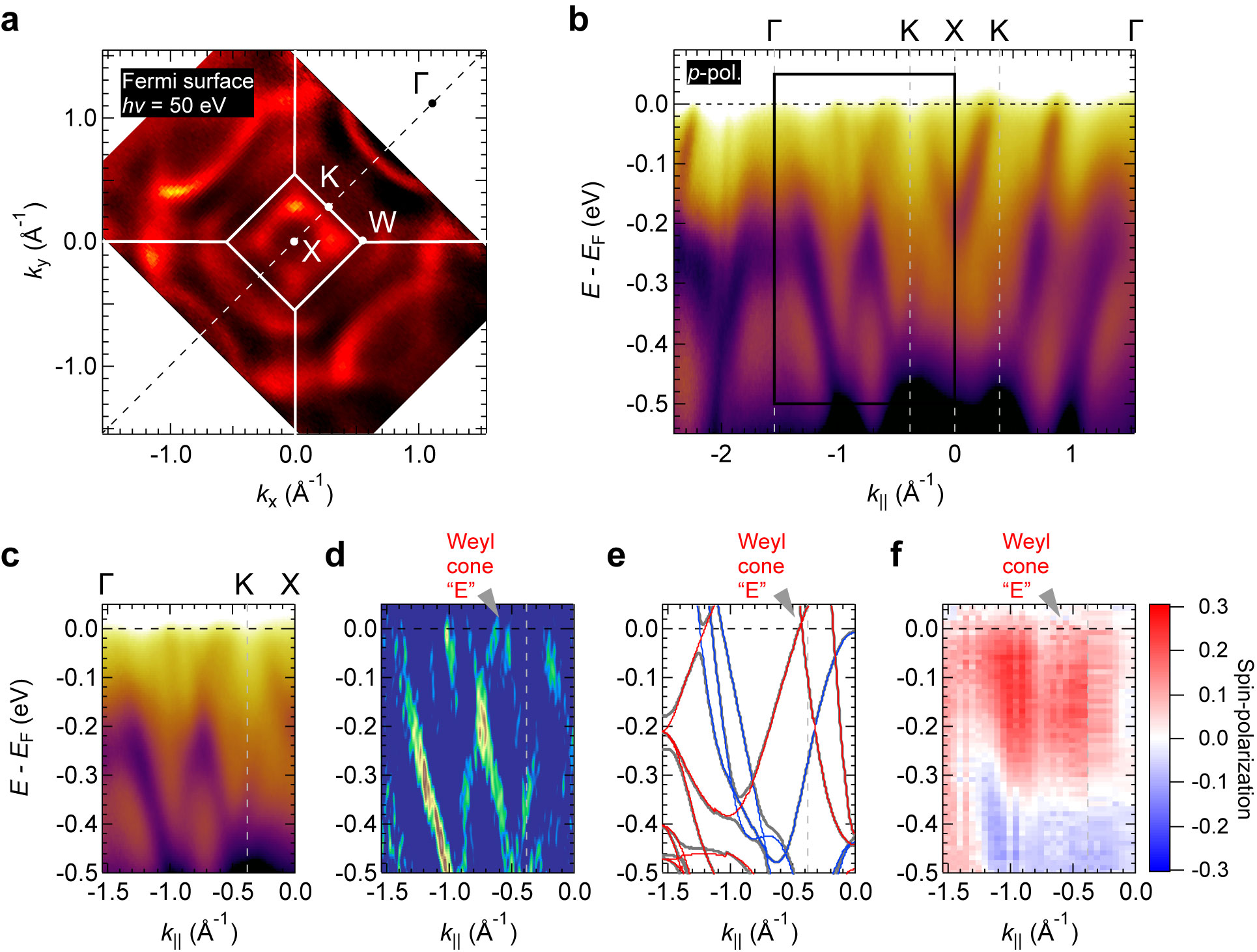}
\caption{\label{fig_hole} \bf {Observed spin-polarized band structures of a hole-doped Co$_2$MnGa film.} \rm(a) Fermi surface of the sample H3* ($N_{\rm v} = 27.3$) recorded at 50 eV with $p$-polarized light. (b) Wide ARPES image along the $\Gamma$--K--X line indicated by the white dashed line in (a). (c, d) Magnified ARPES image in the frame in (b) and its second derivative. (e) Calculated band dispersion with $E_{\rm F}$ shifted downward by 220 meV. (f) Spin-polarization map along the $\Gamma$--K--X line.}
\end{figure*}

In order to compare the band structures of electron-doped and hole-doped samples, we also performed SARPES experiments for the sample H3* ($N_{\rm v} = 27.3$).
Figures \ref{fig_hole}(a) and (b) show the observed Fermi surface and wide range ARPES image along $\Gamma$--K--X line recorded at 50 eV photon energy with $p$-polarized light.
At X point, the electron-pocket crossing $E_{\rm F}$ can be seen while the large hole-pocket is located at $\Gamma$ point.
These features are in good agreement with the electron-doped sample [Fig. \ref{fig_ARPES}].
In Figs. \ref{fig_hole}(c) and (d), we present the magnified ARPES image along the $\Gamma$--K--X direction and its second derivative along the momentum axis.
Around $-$0.60 $\rm \AA^{-1}$ near $E_{\rm F}$, one can see a tilted band, in which the node is slightly located above $E_{\rm F}$ [see gray arrow in Fig. \ref{fig_hole}(d)].
As we can see in Fig. \ref{fig_hole}(f), this tilted band has a strong majority-spin component.
These experimental results agree with the results of the calculation, shown in Fig. \ref{fig_hole}(e) with the position of $E_{\rm F}$ shifted downward by 220 meV. 
Therefore, we conclude that the tilted cone close to K point belongs to the Weyl cone E [see Fig. 2(a)].

Finally, we discuss the role of the Fermi energy tuning in optimizing the anomalous transport properties. From ARPES measurements from two samples, we determined the relative $E_{\rm F}$ shifts from the stoichiometric condition ($N_{\rm v}=28.0$), specifically, +70 meV for the sample E2 ($N_{\rm v}=28.5$) and $-220$ meV for the sample H3* ($N_{\rm v}=27.3$). The latter has lower $\sigma^{\rm A}_{xy}$ and $\alpha^{\rm A}_{xy}$ than the former. These results reasonably explain the observed and calculated large anomalous transport properties presented in Fig. \ref{fig_trans} based on our calculations given in Fig. \ref{fig_theory}. In detail, because $E_{\rm F}$ of sample E2 is closely located at the gapped region of the Weyl cones A and C through slight electron doping, it satisfies a criterion whereby a large Berry curvature in both $k_z=0$ and $k_z=2\pi/a$ planes is generated and thereby results in a gigantic ANE because the $\alpha^{\rm A}_{xy}$ is large. Therefore, we have clearly confirmed that enhancing the anomalous transport properties of topological ferromagnets involves two key features: (1) the creation of a massive (gapped) Weyl cone from the nodal loop by mirror-symmetry breaking and (2) the tuning of the Fermi energy.

\section{Conclusion}
In summary, we have experimentally and theoretically investigated a one-to-one correspondence between the electronic structure and the anomalous transport properties in the film form of Co$_2$MnGa topological ferromagnets. By {\it in-situ} SARPES, we provided a direct visualization of the spin-polarized and massive Weyl cones and the peculiar surface state under mirror-symmetry breaking. When $E_{\rm F}$ approaches the gap of the massive Weyl cones by the electron doping, we have recorded the highest anomalous Nernst thermopower (6.2 $\rm \mu V K^{-1}$) at room temperature among ferromagnetic films to the best of our knowledge. Our findings signify the insufficient $E_{\rm F}$ position tuning against the Weyl cone is the most probable cause for the smaller anomalous Nernst thermopower in the Co$_2$MnGa films reported in previous studies~\cite{2018Reichlova-APL, 2020Park-PRB} and provide the reliable guiding principle to maximize the Nernst thermopower by the band engineering utilizing the SARPES, transport measurements, and ab-initio calculations, for the first time. From an applications perspective, our work facilitates the implementation of various novel applications based on the thin film form of topological magnets; namely, the large transverse electric and thermoelectric conversions without an external magnetic field, which cannot be realized in the bulk form because of the formation of magnetic domains, are promising for novel thermoelectric applications such as heat flux sensor.

\section*{Methods}
\subsection*{Thin film growth}
Epitaxial thin films of (001)-oriented Co$_2$MnGa having different compositions were deposited on a MgO(001) single crystalline substrate at 600$^\circ$C using a co-sputtering technique with Co, Mn, and Co$_{41.2}$Mn$_{27.5}$Ga$_{31.3}$ sputtering targets. The films were designed to be 50 nm thick. To prevent the film from oxidizing for transport measurements, a 1.5 nm-thick Al capping layer was deposited by rf magnetron sputtering. The base pressure of the deposition chamber was near $2 \times 10^{-7}$ Pa. The thickness and the composition of the film were evaluated by wavelength dispersive X-ray fluorescence analysis. Table \ref{table} shows the result of the composition analysis and the evaluated total valence electron number for all prepared Co$_2$MnGa films.

For SARPES measurements, uncapped films were deposited on a MgO substrate with a buffer layers of Cr (10 nm) and Ag (100 nm) to smooth the surface. To avoid surface contaminations, grown films were transferred from the magnetron sputtering chamber to the preparation chamber of the SARPES instrument using a portable suitcase chamber to avoid exposure to air ($<$ $1 \times 10^{-6}$ Pa). Note that sample H3* was deposited on the MgO(001) substrate with buffer layers at room temperature, and then post-annealed at 550$^\circ$C for 30 minutes.

\begin{table}[htb]
  \caption{\label{table} \bf {Atomic compositions and valence electron numbers of Co$_2$MnGa films.} 
  \rm Results of the composition analysis using XRF and estimated valence electron number ($N_{\rm v}$) for all eleven Co$_2$MnGa films prepared.}
  \begin{tabular}{|c|c|c|c|c|} \hline
    Sample name & Co(at.\%) & Mn(at.\%) & Ga(at.\%) & $N_{\rm v}$ \\ \hline\hline
    E2 & 53.0 & 23.8 & 23.2 & 28.5 \\
    E1 & 52.6 & 22.3 & 25.1 & 28.2 \\
    Stoichiometry & 50.0 & 25.0 & 25.0 & 28.0 \\
    H1 & 47.4 & 25.1 & 27.5 & 27.4 \\
    H2 & 47.6 & 24.6 & 27.8 & 27.4 \\
    H3* & 45.3 & 27.6 & 27.1 & 27.3 \\
    H4 & 44.9 & 27.0 & 28.1 & 27.1 \\
    H5 & 45.6 & 24.9 & 29.5 & 26.9 \\
    H6 & 41.4 & 27.9 & 30.7 & 26.4 \\
    H7 & 38.5 & 29.3 & 32.2 & 25.9 \\
    H8 & 35.4 & 31.8 & 32.8 & 25.6 \\  \hline
  \end{tabular}
\end{table}

\subsection*{Measurement of transport and magnetic properties}
The magnetic properties of the Co$_2$MnGa films were measured with a superconducting quantum interference device-vibrating sample magnetometer (SQUID-VSM, Quantum Design Co. Ltd). The crystal structure was revealed through X-ray diffraction with a Cu K$\rm \alpha$ X-ray source and a two-dimensional detector (PILATUS 100K/R, Rigaku Co.). The substrate was cleaved to a size with lateral dimensions of $\sim$7.0 $\times$ 10.0 mm$^2$, and then the film was patterned into a Hall bar structure with a width of 2.0 mm and a length of 7.0 mm through photolithography and Ar ion milling. The AHE was measured at 300 K applying a 1 mA electric current along the [110] direction and a magnetic field was applied perpendicular to the [001] direction of the Co$_{2}$MnGa film using a Physical Property Measurement System (PPMS, Quantum Design Co., Ltd.). The ANE was also measured at 300 K applying a temperature gradient $\nabla T_{\rm in}$ along the [110] direction and a magnetic field along the [001] direction of the Co$_2$MnGa film in PPMS. To evaluate $\nabla T_{\rm in}$, the Seebeck coefficient $S_{xx}$ in the Co$_{2}$MnGa film was obtained outside the PPMS by measuring the $\nabla T_{\rm out}$ using an infrared thermal camera and a black-body coating to calibrate the emissivity. Then, $\nabla T_{\rm in}$ can be estimated from the Seebeck voltage obtained in the PPMS and $S_{xx}$. The accuracy of this method to evaluate ANE has been confirmed in the previous studies~\cite{2019Nakayama-PRM, 2020Sakuraba-PRB}.

\subsection*{Theoretical calculations}
We calculated the electronic structure of $L2_1$-ordered Co$_2$MnGa applying the full-potential linearized augmented plane-wave method including SOI, which is implemented in the WIEN2k program \cite{wien2k}. The lattice constant of the cubic unit cell was set to the experimentally determined value of 5.755\,{\AA} and the $k$-point number in the self-consistent-field calculation was chosen as 20\,$\times$\,20\,$\times$\,20 after confirming the convergence of the total energy. Using the electronic structure obtained, we calculated the anomalous Hall conductivity $\sigma^{\rm A}_{xy}$ using~\cite{2004Yao-PRL}
\begin{eqnarray}
\sigma^{\rm A}_{xy}(\epsilon)&=&-\frac{e^2}{\hbar}
\int \frac{d^3k}{(2\pi)^3}\,\, \Omega^{z}({\bf k}),\label{eq:sigma}\\
\Omega^{z}({\bf k})&=&- {\left(\frac{\hbar}{m}\right)}^2\,
\sum_{n} f(E_{n,{\bf k}},\epsilon)
\sum_{n' \neq n} \frac{2\,{\rm Im} \langle \psi_{n,{\bf k}} |p_x| \psi_{n',{\bf k}} \rangle
\langle \psi_{n',{\bf k}} |p_y| \psi_{n,{\bf k}} \rangle}{(E_{n',{\bf k}}-E_{n,{\bf k}})^2},
\end{eqnarray}
where $n$ and $n'$ denote band indices, $\Omega^{z}({\bf k})$ denotes the Berry curvature, $p_x$ ($p_y$) the $x$ ($y$) component of the momentum operator, $\psi_{n,{\bf k}}$ the eigenstate with the eigenenergy $E_{n,{\bf k}}$, and $f(E_{n,{\bf k}},\epsilon)$ the Fermi distribution function for the band $n$ and the wave vector ${\bf k}$ at the energy $\epsilon$ relative to the Fermi energy. In the calculation of $\sigma^{\rm A}_{xy}$, the direction of the magnetization was set along the [100] direction [Fig. \ref{fig_trans}(a)] and 90\,$\times$\,90\,$\times$\,90 $k$ points were used for the Brillouin zone integration ensuring good convergence for $\sigma^{\rm A}_{xy}$.

From Boltzmann transport theory, we calculated the transverse thermoelectric conductivity $\alpha^{\rm A}_{xy}$ for a given temperature $T$ by substituting the obtained $\sigma^{\rm A}_{xy}$ into the following expression:
\begin{equation}
\alpha^{\rm A}_{xy}=-\frac{1}{eT}\int \,d\epsilon \left(-\frac{\partial f}{\partial \epsilon} \right) (\epsilon-\mu)\, \sigma^{\rm A}_{xy}(\epsilon),\label{eq_alphaxy}
\end{equation}
where $f=1/[\exp((\epsilon-\mu)/k_{\rm B}T)+1]$ denotes the Fermi distribution function with $\mu$ the chemical potential. Here, $\mu=0$ corresponds to the Fermi level.

\subsection*{Spin- and angle-resolved photoelectron spectroscopy}
ARPES and SARPES measurements were performed at the ESPRESSO end-station (BL-9B) in the Hiroshima Synchrotron Radiation Center~\cite{2013Okuda-RSI, 2015Okuda-JESRP}. The base pressure of the SARPES chamber was $5 \times 10^{-9}$ Pa. The photoelectrons were acquired using a hemispherical electron analyser (R4000, Scienta-Omicron). The spin-polarization was measured using a very low-energy electron diffraction-type spin detector. The experimental geometry is shown in Supplementary Fig. 4. The energy and angular resolutions for ARPES (SARPES) were set to 45 meV (55 meV) and $\pm$0.3$^\circ$ ($\pm$1.5$^\circ$), respectively. The effective Sherman function was 0.28 for the SARPES measurements.

During all SARPES measurements, the temperature was maintained below 40 K. Before each ARPES and SARPES measurement, the sample was annealed at 550 $^\circ$C for 30 minutes at the preparation chamber (base pressure $\sim$ $4 \times 10^{-8}$ Pa). The quality and cleanliness of the annealed sample was checked by low-energy electron diffraction (Supplementary Fig. 3 and the associated text). A magnetic field as large as $\sim$0.1 T was applied to the samples along the [100] for sample E2 ([110] for sample H3*) easy-axis using a permanent magnet in the preparation chamber at room temperature. A 0.1 T magnetic field was sufficiently high to saturate the magnetization (Supplementary Fig. 2 and the associated text). The ratio between the remanent and saturation magnetizations along the [100] easy-axis of the sample E2 is 0.96. Therefore, a single magnetic domain was overall obtained for SARPES measurements.

\section*{Data Availability}
The data presented in this paper are available from the authors on reasonable request.

\section*{Acknowledgements}
The SARPES measurements were performed with the approval of the Proposal Assessing Committee of Hiroshima Synchrotron Radiation Center (Proposals Nos.\ 18BG038 and 19AG054). This work was financially supported by KAKENHI (Nos.\ 16H02114, 17H06152, 17H06138, and 18H03683). K.S. was financially supported by a Grant-in-Aid for JSPS Fellows (No.\ 19J00858). Y.S. was financially supported by a JSPS KAKENHI Grant-in-Aid for Young Scientists (A) (No.\ JP2670945) and PRESTO from the Japan Science and Technology Agency (No.\ JPMJPR17R5). We thank S. Kurdi, A. Sakuma, K. Nawa, K. Uchida, and K. Hono for valuable discussions and N. Kojima and B. Masaoka for a technical support.

\section*{Author contributions}
K.S., T.K. and M.K. performed the SARPES experiments with the assistance of K.Mi and T.O.
Y.S., K.G. and W.Z. synthesized the thin films and performed the transport measurements. K.Ma and Y.M. performed the {\it ab initio} calculations. K.S., Y.S. and K.Ma wrote the manuscript with inputs from all authors. A.K. supervised the work.

\section*{Competing interests}
The authors declare no competing interests.
\end{document}